\newcommand\IPA{IPA-CuCl$_3$}
\documentclass[twocolumn,prb,showpacs,preprintnumbers,amsmath,amssymb]{revtex4}

\usepackage{graphicx}
\usepackage{dcolumn}
\usepackage{bm}

\begin{document}

\title{Effect of pressure on the quantum spin ladder material \IPA.}

\author{Tao Hong}
\author{V. O. Garlea}
\author{A. Zheludev}
\author{J. Fernandez-Baca}
\affiliation{Neutron Scattering Sciences Division, Oak Ridge
National Laboratory, Oak Ridge, Tennessee 37831-6393, USA.}

\author{H. Manaka}
\affiliation{Graduate School of Science and Engineering, Kagoshima
University, Korimoto, Kagoshima 890-0065, Japan.}

\author{S. Chang}
\author{J. B. Leao}
\author{S. J. Poulton}
\affiliation{ NIST Center for Neutron Research, National Institute
of Standards and Technology, Gaithersburg, MD 20899, USA.}

\date{\today}

\begin{abstract}
Inelastic neutron scattering and bulk magnetic susceptibility
studies of the quantum S=1/2 spin ladder system \IPA\ are
performed under hydrostatic pressure. The pressure dependence of
the spin gap $\Delta$ is determined. At $P=1.5$~GPa it is reduced to
$\Delta=0.79$~meV from $\Delta=1.17$~meV at ambient pressure. The
results allow us to predict a soft-mode quantum phase transition
in this system at P$_\mathrm{c}\sim 4$~GPa. The measurements are
complicated by a proximity of a structural phase transition that
leads to a deterioration of the sample.
\end{abstract}


\maketitle

\section{Introduction}
Phase transitions in quantum spin liquids have recently been
attracting a great deal of attention.\cite{Sachdev2008} Such
systems remain disordered at $T=0$ due to zero point quantum spin
fluctuations and have an energy gap $\Delta$ in the magnetic
excitation spectrum. If changing some external parameter, such as
magnetic field or hydrostatic pressure, reduces the gap energy, a
soft-mode quantum phase transition can be expected at the point
where $\Delta\rightarrow 0$. Beyond the phase transition the system
typically develops long-range magnetic order in the ground state.
Field-induced ordering transitions are the easiest to realize
experimentally and the most extensively
studied.\cite{Shiramura1997,Kodama2002,Ruegg2003,Sebastian2005,Garlea2007}
In that scenario one of the three lowest energy $S=1$ gap
excitations is driven to zero energy by virtue of Zeeman effect.
In a Heisenberg spin gap system such a transition breaks $SO(2)$
symmetry and is famously described as a Bose-Einstein condensation
of magnons (BEC).\cite{Giamarchi2008}

Less studied are transitions driven by external pressure. These
may occur in materials in which pressure-induced lattice
distortions modify the exchange constants or magnetic anisotropy
in such a way, as to reduce the gap energy.\cite{Schmeltzer1998}
The transition is distinct from BEC in that it leads to a
spontaneous violation of $SO(3)$ symmetry in a Heisenberg system.
To date, only one experimental realization of this latter
mechanism has been found, namely in the $S=1/2$ spin-dimer system
TlCuCl$_3$.\cite{Ruegg2004,Ruegg2008} This compound is extremely
sensitive to the effect of pressure, and the transition is
observed at a critical value as low as $P_\mathrm{c}=107$~MPa.

The present work focuses on another prototypical quantum spin
liquid, namely the $S=1/2$ quantum spin ladder material
\IPA.\cite{Manaka97,Manaka98,Masuda2006} The crystal structure and
topology of magnetic interactions in this Heisenberg
antiferromagnet were discussed in detail in
Ref.~\onlinecite{Masuda2006}. Cu$^{2+}$-based $S=1/2$ ladders with
antiferromagnetic leg coupling run along the \textbf{a} axis of the
triclinic $P\overline{1}$ crystal structure. Spin correlations along the
ladder rungs are ferromagnetic. Additional non-frustrating AF
interactions are along the ladder diagonals. The ground state is a
spin singlet with a spin gap $\Delta=1.17$~meV. Small but
measurable interactions between ladders account for a weak
dispersion of gap excitations along the crystallographic \textbf{c}
axis. Interactions along the \textbf{b} axis are negligible, as is
magnetic anisotropy. The global minimum of the 3D dispersion is
located at the magnetic zone-center $(0.5,0,0)$. It is at this
point where a magnetic Bragg peak appears when BEC of magnons and
long-range ordering are induced in \IPA\ by an external field
exceeding $H_c=9.7$~T.\cite{Garlea2007,Zheludev2007} The main
purpose of the present experiments is an attempt to suppress the
gap and potentially induce an ordering transition in \IPA\ by
applying hydrostatic pressure, rather than a magnetic field.

\begin{figure}
 \includegraphics[width=1.0\columnwidth]{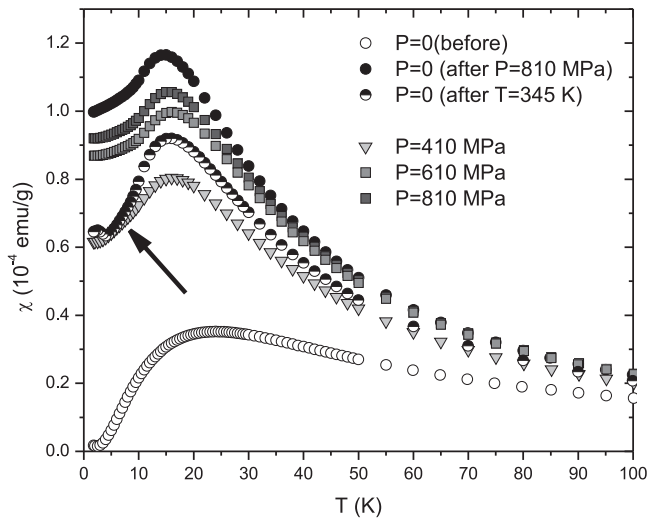}
 \caption{Temperature dependence of the magnetic susceptibility in \IPA\ is measured at various pressures in a DC field of 1 T. Samples previously pressurized to 810~MPa (solid circles) show a behavior different from that of never pressurized ones (open circles), but similar to that of those that have gone a structural transition at $T=323$~K at ambient pressure (half-filled circles). The error bars are smaller than the data points. The arrow indicates a non-activated contribution at $T\rightarrow 0$.}
  \label{susc}
\end{figure}

\section{Bulk measurements}
While less straightforward to interpret, bulk magnetic
measurements are typically much easier to perform under pressure
than neutron scattering experiments. For \IPA\ we investigated the
temperature dependence of the magnetic susceptibility down to
$T=1.5$~K, using a superconducting quantum-interference device (SQUID)
magnetometer in a DC field of 1 T and single crystal
samples of a typical size $2\times 6 \times 1$~mm$^3$. Pressures
of up to 810~MPa were achieved using a TiCu piston cylinder clamp cell \cite{Koyama1998} that was loaded at
room temperature. The pressure medium was a mixture of two types of fluid fluorinert (FC70 : FC77 = 1 : 1). The produced pressure around 4.2 K was calibrated as a function of the applied load by means of the Meissner effect of Sn. Sample data are shown in Fig.~\ref{susc}. At
ambient pressure (open circles) the susceptibility curve is
consistent with that reported in Ref.~\onlinecite{Manaka97} and has a
activated character at $T\rightarrow 0$, which is a signature of a
spin gap. At an applied pressure of $P=410$~MPa the
low-temperature behavior changes qualitatively. Rather than
dropping to zero, the $\chi(T)$ tends to a finite value at the lower
end of the measurement range. The trend continues up to 810~MPa,
with the low-temperature value steadily increasing. An unusual
observation is the irreversibility of this effect: releasing the
external pressure {\it does not return the susceptibility to its
original values} (open circles in Fig.~\ref{susc}). Pressurizing
the samples also has an irreversible effect on their visual
appearance. As-grown crystals are dark red/brown, while those
previously pressurized to 810~MPa acquire a pale beige color.

The irreversibility is clearly caused by structural damage to the
samples, which in turn is likely due to a pressure-induced
crystallographic transition. Indeed, \IPA\ is known to be close to
a structural phase boundary. At ambient pressure it undergoes a
structure phase transformation at $T=323$~K.\cite{Roberts1981}
This transition induces the same kind of color change in \IPA\ single crystals as the
application of pressure. The crystal structure at high temperature phase consists of linear chains and three Cu-Cl-Cu superexchange paths.\cite{Roberts1981} The $\chi(T)$ curves for crystals
previously taken through the $T=323$~K transition
(Fig.~\ref{susc}, half-filled circles) are also qualitatively
similar to those collected in previously pressurized samples.

One can expect that the microscopic fracturing induced by the
phase transition will have an effect on the spin ladders in \IPA\
similar to what microfine grinding has on the Haldane spin
chain material NINAZ.\cite{Granroth1998} A fragmentation of the
ladders will release $S=1/2$ degrees of freedom on the ends of
every finite-length fragment. These free spins will constitute the
dominant contribution to susceptibility and specific heat at low
temperatures, where the contribution of uninterrupted ladder
sections is exponentially small. Thus, any effect of the applied
pressure on the spin gap is masked by the appearance of end-chain
spins, and therefore can not be conclusively investigated by bulk
susceptibility or calorimetric techniques. In this case, furthermore, pressure dependence of $\chi(T)$ curves do not show a systematic behavior because the sign as well as the absolute value of the rung exchange interaction is sensitively not only to the Cu-Cl-Cu bonding angle but also to the dihedral folding angle and twisting angle of the Cu$_2$Cl$_6$ dimer planes.\cite{Manaka1997}

\section{Neutron scattering measurements}
The end-chain spins that interfere with bulk measurements have no
intrinsic dynamics and therefore do not severely affect inelastic
neutron experiments. In general, it is difficult to carry out inelastic neutron scattering under high pressure because of the limited sample space and the attenuation of the neutron beam
due to the thick walls needed to contain the high pressure.

The gap excitations in \IPA\ were studied
under pressure in two separate runs on the SPINS 3-axis
cold-neutron spectrometer at NIST. A pyrolytic graphite PG(002) monochromator
was used in conjunction with a flat (Setup 1) or horizontally focusing (Setup 2)
PG analyzer. The incident beam divergence was controlled by the
$^{58}$Ni guide before the monochromator. For setup 1 an 80' Soller collimator was placed
between the sample and analyzer while a 120' radial collimator was used for setup 2. The final energy was fixed at $E_f=5$ meV (Setup 1) or 3.7 meV
(Setup 2) and a Be (Setup 1) or BeO (Setup 2) low-pass filter was placed after the sample.

In the first series of
measurements (Setup 1) we utilized an aluminium He-gas cell that
delivered pressure of up to 650~MPa. This cell has a neutron
transmission of 65\%, and the pressure can be changed {\it in
situ}, without warming the sample up to room temperature, though the \emph{P}-\emph{T} curve for helium
must be considered. Five
fully deuterated single crystals of a total mass $\approx 150$~mg
were co-aligned to an irregular mosaic spread of 2$^\circ$ full
width at half maximum (FWHM) at ambient pressure (each individual
crystal had a mosaic of about 0.5$^\circ$ FWHM). Increasing the
pressure to 620~MPa had no effect on sample mosaic.

\begin{figure}
 \includegraphics[width=1.0\columnwidth]{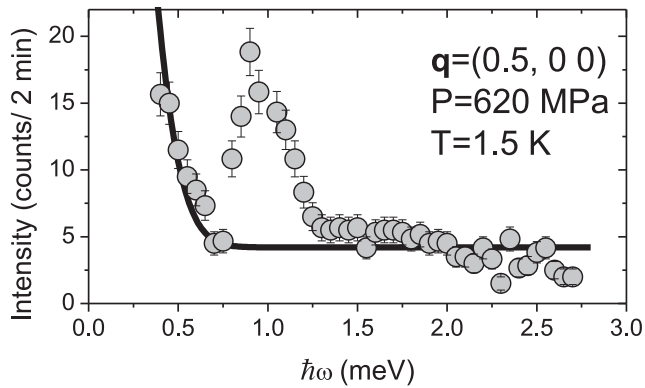}
 \caption{A typical constant-$\textbf{q}$ scan (raw data) collected at the magnetic zone-center in \IPA\
 at $P=620$~MPa using Setup 1 (symbols). The line is a model
 for the background contribution, as described in the text.}
  \label{exdata1}
\end{figure}

\begin{figure}
 \includegraphics[width=1.0\columnwidth]{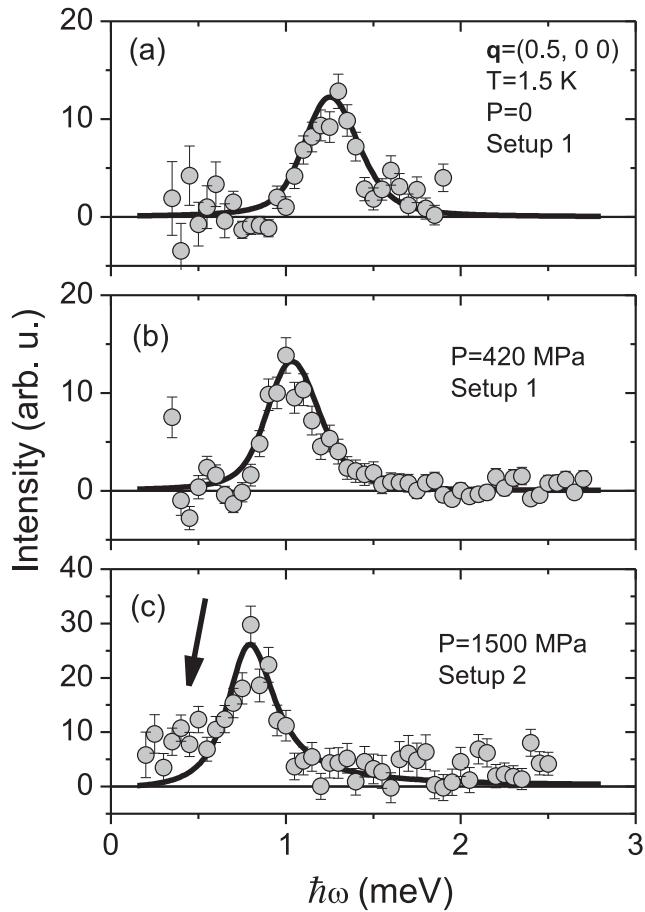}
 \caption{Background-subtracted constant-$q$ scans collected in \IPA\ at the magnetic zone center at various pressures (symbols).
 The solid lines are fits to the data based on a model cross section function convoluted with the spectrometer resolution,
 as described in the text. The arrow indicates a broadening of the inelastic
 peak or extra intensity inside the gap.}
  \label{exdata2}
\end{figure}

The second setup (Setup 2) utilized the Al$_2$O$_3$ clamp-type
pressure cell as described in detail in Ref.~\onlinecite{Onodera1987}. For this cell neutron
transmission is strongly energy-dependent and can be as low as
10\%.\cite{Zaliznyak1998}  Due to the cell design, it can only be loaded
incrementally, each pressure change requiring its removal from the cryostat at room temperature. In the experiment we
utilized a 300~mg deuterated single crystal \IPA\ sample with an
initial mosaic spread of about 0.5$^\circ$. The pressure medium
was fluid fluorinert FC-75. All data were collected at $P=1.5$~GPa where
the mosaic of the crystal was Gaussian in shape, but irreversibly
broadened to as much as 5.5$^\circ$. The actual value of the
applied pressure was determined by measuring the d-spacing in a
single crystal of NaCl loaded in the same cell, following
Ref.~\onlinecite{Onodera1987}. Attempting to pressurize the sample
beyond 1.5~GPa resulted in collapse of the sample. In both experiments the
sample environment was a He-flow cryostat, enabling data
collection at $T=1.5$~K.

All the data were taken in constant-$\textbf{q}$ scans at the magnetic
zone-center $(0.5,0,0)$. The data obtained using Setup 2 were
corrected for the energy dependence of neutron absorbtion using
the results of Ref.~\onlinecite{Zaliznyak1998}. The following
procedure was used to determine the background. Beyond 0.8~meV
energy transfer it was measured away from the magnetic zone-center
at wave vectors $(0.3,0,0)$ and $(0.75,0,0)$, and fit to a
straight line. In the range 0.2-0.8~meV it was taken directly from
the $(0.5,0,0)$ scan at ambient pressure, where no magnetic
scattering is expected due to a 1.2~meV spin gap, and fit to an
additional Gaussian profile to account for elastic incoherent
scattering in the sample environment and thermal diffuse
scattering in the monochromator. Fig.~\ref{exdata1} shows raw data
measured using Setup 1 at 620~MPa (symbols) along with the
estimated background contribution (solid line). Fig.~\ref{exdata2}
shows typical background-subtracted scans obtained using Setups 1
and 2.

\begin{figure}
 \includegraphics[width=1.0\columnwidth]{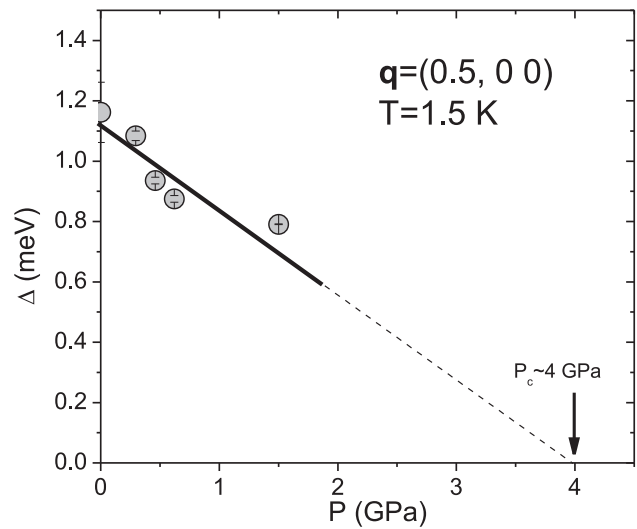}
 \caption{Pressure dependence of the spin  gap in \IPA\ measured using inelastic neutron scattering at $T=1.5$~K.
 A linear extrapolation suggests a soft mode quantum phase transition at $P_c\approx 4$~GPa.}
  \label{gapvsp}
\end{figure}

\section{Data analysis and discussion} The data were analyzed by
least-squares fitting to a parameterized model cross section function
that was numerically convoluted with the calculated resolution of
the spectrometer. We utilized the same two-Lorentzian
representation of the Damped Harmonic Oscillator cross section for
\IPA, as previously employed in the study of finite-temperature
effects on gap excitations.\cite{Zheludev2008} At each pressure,
the parameters of this model are the gap energy $\Delta$, the
excitation width (inverse lifetime) $\Gamma$ and an overall
intensity prefactor:
 \begin{eqnarray}
 S(\mathbf{q},\omega)& \propto &
 (n(\omega)+1)\left[  \frac{\Gamma}{(\omega-\omega_\mathbf{q})^2+\Gamma^2} + \right. \nonumber\\
  & - & \left. \frac{\Gamma}{(\omega+\omega_\mathbf{q})^2+\Gamma^2}
 \right],
 \end{eqnarray}
where $n(\omega)+1$ is the Bose factor and the dispersion relation
$\omega_\mathbf{q}$ is given by Eq.~(2) in
Ref.~\onlinecite{Masuda2006}.

For Setup 1 good fits to the data are obtained in the entire scan
range at all pressures, assuming resolution-limited excitations
with $\Gamma\rightarrow 0$. The results  are plotted in solid
lines in Fig.~\ref{exdata2}a and b. For the 1.5~GPa data set
measured using Setup 2, however, the best fit corresponds to
$\Gamma=0.13 (0.02)$~meV (solid line in Fig.~\ref{exdata2}c). This
intrinsic width primarily accounts for the additional scattering present
at low energies (arrow in Fig.~\ref{exdata2}c). For all experimental pressures
the gap energies extracted from
the fits are plotted vs. pressure in Fig.~\ref{gapvsp}.

Despite the technical challenges, our neutron results established
a steady decrease of the gap energy in \IPA\ under applied
hydrostatic pressure. While it is technically not possible to
reach the quantum critical point in our experiments, a linear
extrapolation of the pressure dependence of $\Delta$ suggests that
$P_c$ will be close to 4~GPa. Such a high pressure is currently
unavailable for inelastic neutron scattering at NCNR. An alternative
technique may be high-field magnetization studies that probe the
gap indirectly, by detecting the {\it field-induced}  quantum
phase transition under applied pressure. In \IPA\ the
corresponding critical field can be expected to decrease, and to
reach zero at the pressure-induced critical point.

The broadening of the inelastic peak at $P=1.5$~GPa in Setup 2 is
to be linked with its drastically increased mosaic spread. It is
most likely due to the fractioning of the crystals upon going
through the structural phase transition discussed previously. As
established theoretically and experimentally, finite lengths of
ladder fragments in a 1D gapped quantum antiferromagnet will
induce a broadening of gap the excitations similar to that at
finite temperature.\cite{Xu2007} Note, that neither the sample
mosaic, nor the inelastic peaks, are affected by pressure in Setup
1. The key difference is that in Setup 1 the sample was never
warmed up above $T\approx 200$~K under pressure, and presumably
never went through a phase transition. Any future studies aiming
at exploring the incipient quantum phase transition in \IPA\
should take heed of this, and aim to pressurize the samples
at low temperature.

Research at ORNL was funded by the United States Department of
Energy, Office of Basic Energy Sciences- Materials Science, under
Contract No. DE-AC05-00OR22725 with UT-Battelle, LLC. The work at
NIST is supported by the National Science Foundation under
Agreement Nos. DMR-9986442, -0086210, and -0454672.


\begin{thebibliography}{18}
\expandafter\ifx\csname
natexlab\endcsname\relax\def\natexlab#1{#1}\fi
\expandafter\ifx\csname bibnamefont\endcsname\relax
  \def\bibnamefont#1{#1}\fi
\expandafter\ifx\csname bibfnamefont\endcsname\relax
  \def\bibfnamefont#1{#1}\fi
\expandafter\ifx\csname citenamefont\endcsname\relax
  \def\citenamefont#1{#1}\fi
\expandafter\ifx\csname url\endcsname\relax
  \def\url#1{\texttt{#1}}\fi
\expandafter\ifx\csname
urlprefix\endcsname\relax\def\urlprefix{URL }\fi
\providecommand{\bibinfo}[2]{#2}
\providecommand{\eprint}[2][]{\url{#2}}

\bibitem[{\citenamefont{Sachdev}(2008)}]{Sachdev2008}
\bibinfo{author}{\bibfnamefont{S.}~\bibnamefont{Sachdev}},
  \bibinfo{journal}{Nature Physics} \textbf{\bibinfo{volume}{4}},
  \bibinfo{pages}{173} (\bibinfo{year}{2008}).

\bibitem[{\citenamefont{Shiramura et~al.}(1997)\citenamefont{Shiramura,
  Takatsu, Tanaka, Kamishima, Takahashi, Mitamura, and Goto}}]{Shiramura1997}
\bibinfo{author}{\bibfnamefont{W.}~\bibnamefont{Shiramura}},
  \bibinfo{author}{\bibfnamefont{K.}~\bibnamefont{Takatsu}},
  \bibinfo{author}{\bibfnamefont{H.}~\bibnamefont{Tanaka}},
  \bibinfo{author}{\bibfnamefont{K.}~\bibnamefont{Kamishima}},
  \bibinfo{author}{\bibfnamefont{M.}~\bibnamefont{Takahashi}},
  \bibinfo{author}{\bibfnamefont{H.}~\bibnamefont{Mitamura}}, \bibnamefont{and}
  \bibinfo{author}{\bibfnamefont{T.}~\bibnamefont{Goto}}, \bibinfo{journal}{J.
  Phys. Soc. Jpn.} \textbf{\bibinfo{volume}{66}}, \bibinfo{pages}{1900}
  (\bibinfo{year}{1997}).

\bibitem[{\citenamefont{Kodama et~al.}(2003)\citenamefont{Kodama, Takigawa,
  Horvatic, Berthier, Kageyama, Ueda, Miyahara, Becca, and Mila}}]{Kodama2002}
\bibinfo{author}{\bibfnamefont{K.}~\bibnamefont{Kodama}},
  \bibinfo{author}{\bibfnamefont{M.}~\bibnamefont{Takigawa}},
  \bibinfo{author}{\bibfnamefont{M.}~\bibnamefont{Horvatic}},
  \bibinfo{author}{\bibfnamefont{C.}~\bibnamefont{Berthier}},
  \bibinfo{author}{\bibfnamefont{H.}~\bibnamefont{Kageyama}},
  \bibinfo{author}{\bibfnamefont{Y.}~\bibnamefont{Ueda}},
  \bibinfo{author}{\bibfnamefont{S.}~\bibnamefont{Miyahara}},
  \bibinfo{author}{\bibfnamefont{F.}~\bibnamefont{Becca}}, \bibnamefont{and}
  \bibinfo{author}{\bibfnamefont{F.}~\bibnamefont{Mila}},
  \bibinfo{journal}{Science} \textbf{\bibinfo{volume}{298}},
  \bibinfo{pages}{395} (\bibinfo{year}{2003}).

\bibitem[{\citenamefont{Ruegg et~al.}(2003)\citenamefont{Ruegg, Cavadini,
  Furrer, Gudel, Kramer, Mutka, Wildes, Habicht, and Vorderwisch}}]{Ruegg2003}
\bibinfo{author}{\bibfnamefont{C.}~\bibnamefont{Ruegg}},
  \bibinfo{author}{\bibfnamefont{N.}~\bibnamefont{Cavadini}},
  \bibinfo{author}{\bibfnamefont{A.}~\bibnamefont{Furrer}},
  \bibinfo{author}{\bibfnamefont{H.-U.} \bibnamefont{Gudel}},
  \bibinfo{author}{\bibfnamefont{K.}~\bibnamefont{Kramer}},
  \bibinfo{author}{\bibfnamefont{H.}~\bibnamefont{Mutka}},
  \bibinfo{author}{\bibfnamefont{A.}~\bibnamefont{Wildes}},
  \bibinfo{author}{\bibfnamefont{K.}~\bibnamefont{Habicht}}, \bibnamefont{and}
  \bibinfo{author}{\bibfnamefont{P.}~\bibnamefont{Vorderwisch}},
  \bibinfo{journal}{Nature} \textbf{\bibinfo{volume}{423}}, \bibinfo{pages}{62}
  (\bibinfo{year}{2003}).

\bibitem[{\citenamefont{Sebastian et~al.}(2005)\citenamefont{Sebastian, Sharma,
  Jaime, Harrison, Correa, Balicas, Kawashima, Batista, and
  Fisher}}]{Sebastian2005}
\bibinfo{author}{\bibfnamefont{S.~E.} \bibnamefont{Sebastian}},
  \bibinfo{author}{\bibfnamefont{P.~A.} \bibnamefont{Sharma}},
  \bibinfo{author}{\bibfnamefont{M.}~\bibnamefont{Jaime}},
  \bibinfo{author}{\bibfnamefont{N.}~\bibnamefont{Harrison}},
  \bibinfo{author}{\bibfnamefont{V.}~\bibnamefont{Correa}},
  \bibinfo{author}{\bibfnamefont{L.}~\bibnamefont{Balicas}},
  \bibinfo{author}{\bibfnamefont{N.}~\bibnamefont{Kawashima}},
  \bibinfo{author}{\bibfnamefont{C.~D.} \bibnamefont{Batista}},
  \bibnamefont{and} \bibinfo{author}{\bibfnamefont{I.~R.}
  \bibnamefont{Fisher}}, \bibinfo{journal}{Phys. Rev. B}
  \textbf{\bibinfo{volume}{72}}, \bibinfo{pages}{100404(R)}
  (\bibinfo{year}{2005}).

\bibitem[{\citenamefont{Garlea et~al.}(2007)\citenamefont{Garlea, Zheludev,
  Masuda, Manaka, Regnault, Ressouche, Grenier, Chung, Qiu, Habicht
  et~al.}}]{Garlea2007}
\bibinfo{author}{\bibfnamefont{V.~O.} \bibnamefont{Garlea}},
  \bibinfo{author}{\bibfnamefont{A.}~\bibnamefont{Zheludev}},
  \bibinfo{author}{\bibfnamefont{T.}~\bibnamefont{Masuda}},
  \bibinfo{author}{\bibfnamefont{H.}~\bibnamefont{Manaka}},
  \bibinfo{author}{\bibfnamefont{L.-P.} \bibnamefont{Regnault}},
  \bibinfo{author}{\bibfnamefont{E.}~\bibnamefont{Ressouche}},
  \bibinfo{author}{\bibfnamefont{B.}~\bibnamefont{Grenier}},
  \bibinfo{author}{\bibfnamefont{J.-H.} \bibnamefont{Chung}},
  \bibinfo{author}{\bibfnamefont{Y.}~\bibnamefont{Qiu}},
  \bibinfo{author}{\bibfnamefont{K.}~\bibnamefont{Habicht}},
  \bibinfo{author}{\bibfnamefont{K.}~\bibnamefont{Kiefer}},
  \bibnamefont{and}
  \bibinfo{author}{\bibfnamefont{M.}~\bibnamefont{Boehm}},
  \bibinfo{journal}{Phys. Rev. Lett.}
  \textbf{\bibinfo{volume}{98}}, \bibinfo{pages}{167202}
  (\bibinfo{year}{2007}).

\bibitem[{\citenamefont{Giamarchi et~al.}(2008)\citenamefont{Giamarchi, Ruegg,
  and Tchernyshev}}]{Giamarchi2008}
\bibinfo{author}{\bibfnamefont{T.}~\bibnamefont{Giamarchi}},
  \bibinfo{author}{\bibfnamefont{C.}~\bibnamefont{Ruegg}}, \bibnamefont{and}
  \bibinfo{author}{\bibfnamefont{O.}~\bibnamefont{Tchernyshev}},
  \bibinfo{journal}{Nature Physics} \textbf{\bibinfo{volume}{4}},
  \bibinfo{pages}{198} (\bibinfo{year}{2008}).

\bibitem{Schmeltzer1998} D. Schmeltzer and A. R. Bishop, Phys.
Rev. {\bf 58}, R5905 (1998).



\bibitem[{\citenamefont{Ruegg et~al.}(2004)\citenamefont{Ruegg, Furrer,
  Sheptyakov, Strassle, Kramer, Gudel, and Melesi}}]{Ruegg2004}
\bibinfo{author}{\bibfnamefont{C.}~\bibnamefont{Ruegg}},
  \bibinfo{author}{\bibfnamefont{A.}~\bibnamefont{Furrer}},
  \bibinfo{author}{\bibfnamefont{D.}~\bibnamefont{Sheptyakov}},
  \bibinfo{author}{\bibfnamefont{T.}~\bibnamefont{Strassle}},
  \bibinfo{author}{\bibfnamefont{K.~W.} \bibnamefont{Kramer}},
  \bibinfo{author}{\bibfnamefont{H.-U.} \bibnamefont{Gudel}}, \bibnamefont{and}
  \bibinfo{author}{\bibfnamefont{L.}~\bibnamefont{Melesi}},
  \bibinfo{journal}{Phys. Rev. Lett.} \textbf{\bibinfo{volume}{93}},
  \bibinfo{pages}{257201} (\bibinfo{year}{2004}).

\bibitem[{\citenamefont{Ruegg et~al.}(2008)\citenamefont{Ruegg, Normand,
  Matsumoto, Furrer, McMorrow, Kramer, Gudel, Gvasaliya, Mutka, and
  Boehm}}]{Ruegg2008}
\bibinfo{author}{\bibfnamefont{C.}~\bibnamefont{Ruegg}},
  \bibinfo{author}{\bibfnamefont{B.}~\bibnamefont{Normand}},
  \bibinfo{author}{\bibfnamefont{M.}~\bibnamefont{Matsumoto}},
  \bibinfo{author}{\bibfnamefont{A.}~\bibnamefont{Furrer}},
  \bibinfo{author}{\bibfnamefont{D.~F.}~\bibnamefont{McMorrow}},
  \bibinfo{author}{\bibfnamefont{K.~W.} \bibnamefont{Kramer}},
  \bibinfo{author}{\bibfnamefont{H.-U.} \bibnamefont{Gudel}},
  \bibinfo{author}{\bibfnamefont{S.~N.} \bibnamefont{Gvasaliya}},
  \bibinfo{author}{\bibfnamefont{H.}~\bibnamefont{Mutka}}, \bibnamefont{and}
  \bibinfo{author}{\bibfnamefont{M.}~\bibnamefont{Boehm}},
  \bibinfo{journal}{Phys. Rev. Lett.} \textbf{\bibinfo{volume}{100}},
  \bibinfo{pages}{205701} (\bibinfo{year}{2008}).

\bibitem[{\citenamefont{Masuda et~al.}(2006)\citenamefont{Masuda, Zheludev,
  Manaka, Regnault, Chung, and Qiu}}]{Masuda2006}
\bibinfo{author}{\bibfnamefont{T.}~\bibnamefont{Masuda}},
  \bibinfo{author}{\bibfnamefont{A.}~\bibnamefont{Zheludev}},
  \bibinfo{author}{\bibfnamefont{H.}~\bibnamefont{Manaka}},
  \bibinfo{author}{\bibfnamefont{L.-P.} \bibnamefont{Regnault}},
  \bibinfo{author}{\bibfnamefont{J.-H.} \bibnamefont{Chung}}, \bibnamefont{and}
  \bibinfo{author}{\bibfnamefont{Y.}~\bibnamefont{Qiu}},
  \bibinfo{journal}{Phys. Rev. Lett.} \textbf{\bibinfo{volume}{96}},
  \bibinfo{pages}{047210} (\bibinfo{year}{2006}).

\bibitem[{\citenamefont{Manaka et~al.}(1997)\citenamefont{Manaka, Yamada, and
  Yamaguchi}}]{Manaka97}
\bibinfo{author}{\bibfnamefont{H.}~\bibnamefont{Manaka}},
  \bibinfo{author}{\bibfnamefont{I.}~\bibnamefont{Yamada}}, \bibnamefont{and}
  \bibinfo{author}{\bibfnamefont{K.}~\bibnamefont{Yamaguchi}},
  \bibinfo{journal}{J. Phys. Soc. Jpn.} \textbf{\bibinfo{volume}{66}},
  \bibinfo{pages}{564} (\bibinfo{year}{1997}).

\bibitem[{\citenamefont{Manaka et~al.}(1998)\citenamefont{Manaka, Yamada,
  Honda, Katori, and Katsumata}}]{Manaka98}
\bibinfo{author}{\bibfnamefont{H.}~\bibnamefont{Manaka}},
  \bibinfo{author}{\bibfnamefont{I.}~\bibnamefont{Yamada}},
  \bibinfo{author}{\bibfnamefont{Z.}~\bibnamefont{Honda}},
  \bibinfo{author}{\bibfnamefont{H.~A.} \bibnamefont{Katori}},
  \bibnamefont{and}
  \bibinfo{author}{\bibfnamefont{K.}~\bibnamefont{Katsumata}},
  \bibinfo{journal}{J. Phys. Soc. Jpn.} \textbf{\bibinfo{volume}{67}},
  \bibinfo{pages}{3913} (\bibinfo{year}{1998}).

\bibitem[{\citenamefont{Zheludev et~al.}(2007)\citenamefont{Zheludev, Garlea,
  Masuda, Manaka, Regnault, Ressouche, Grenier, Chung, Qiu, Habicht
  et~al.}}]{Zheludev2007}
\bibinfo{author}{\bibfnamefont{A.}~\bibnamefont{Zheludev}},
  \bibinfo{author}{\bibfnamefont{V.~O.} \bibnamefont{Garlea}},
  \bibinfo{author}{\bibfnamefont{T.}~\bibnamefont{Masuda}},
  \bibinfo{author}{\bibfnamefont{H.}~\bibnamefont{Manaka}},
  \bibinfo{author}{\bibfnamefont{L.-P.} \bibnamefont{Regnault}},
  \bibinfo{author}{\bibfnamefont{E.}~\bibnamefont{Ressouche}},
  \bibinfo{author}{\bibfnamefont{B.}~\bibnamefont{Grenier}},
  \bibinfo{author}{\bibfnamefont{J.-H.} \bibnamefont{Chung}},
  \bibinfo{author}{\bibfnamefont{Y.}~\bibnamefont{Qiu}},
  \bibinfo{author}{\bibfnamefont{K.}~\bibnamefont{Habicht}},
  \bibinfo{author}{\bibfnamefont{K.}~\bibnamefont{Kiefer}},
  \bibnamefont{and}
  \bibinfo{author}{\bibfnamefont{M.}~\bibnamefont{Boehm}},
  \bibinfo{journal}{Phys. Rev. B}
  \textbf{\bibinfo{volume}{76}}, \bibinfo{pages}{054450}
  (\bibinfo{year}{2007}).

\bibitem[{\citenamefont{Koyama et~al.}(1998)\citenamefont{Koyama, Goto,
  Kanomata,and Note}}]{Koyama1998}
\bibinfo{author}{\bibfnamefont{K.}~\bibnamefont{Koyama}},
  \bibinfo{author}{\bibfnamefont{T.}~\bibnamefont{Goto}},
  \bibinfo{author}{\bibfnamefont{T.}~\bibnamefont{Kanomata}}, \bibnamefont{and}
  \bibinfo{author}{\bibfnamefont{R.}~\bibnamefont{Note}},
  \bibinfo{journal}{J. Phys. Soc. Jpn.}
  \textbf{\bibinfo{volume}{68}}, \bibinfo{pages}{1693} (\bibinfo{year}{1998}).

\bibitem[{\citenamefont{Roberts et~al.}(1981)\citenamefont{Robert, Bloomquist,
  Willett,and Dodgen}}]{Roberts1981}
\bibinfo{author}{\bibfnamefont{S.~A.} \bibnamefont{Roberts}},
  \bibinfo{author}{\bibfnamefont{D.~R.}~\bibnamefont{Bloomquist}},
  \bibinfo{author}{\bibfnamefont{R.~D.} \bibnamefont{Willett}}, \bibnamefont{and}
  \bibinfo{author}{\bibfnamefont{H.~W.}~\bibnamefont{Dodgen}},
  \bibinfo{journal}{J. Am. Chem. Soc.}
  \textbf{\bibinfo{volume}{103}}, \bibinfo{pages}{2603} (\bibinfo{year}{1981}).


\bibitem[{\citenamefont{Granroth et~al.}(1998)\citenamefont{Granroth, Maegawa,
  Meisel, Krzystek, Brunel, Bell, Adair, Ward, Fanucci, Chou
  et~al.}}]{Granroth1998}
\bibinfo{author}{\bibfnamefont{G.~E.} \bibnamefont{Granroth}},
  \bibinfo{author}{\bibfnamefont{S.}~\bibnamefont{Maegawa}},
  \bibinfo{author}{\bibfnamefont{M.~W.} \bibnamefont{Meisel}},
  \bibinfo{author}{\bibfnamefont{J.}~\bibnamefont{Krzystek}},
  \bibinfo{author}{\bibfnamefont{L.-C.} \bibnamefont{Brunel}},
  \bibinfo{author}{\bibfnamefont{N.~S.} \bibnamefont{Bell}},
  \bibinfo{author}{\bibfnamefont{J.~H.} \bibnamefont{Adair}},
  \bibinfo{author}{\bibfnamefont{B.~H.} \bibnamefont{Ward}},
  \bibinfo{author}{\bibfnamefont{G.~E.} \bibnamefont{Fanucci}},
  \bibinfo{author}{\bibfnamefont{L.-K.} \bibnamefont{Chou}},
  \bibnamefont{and}
  \bibinfo{author}{\bibfnamefont{D.~R.} \bibnamefont{Talham}},
  \bibinfo{journal}{Phys. Rev. B}
  \textbf{\bibinfo{volume}{58}}, \bibinfo{pages}{9312} (\bibinfo{year}{1998}).

\bibitem[{\citenamefont{Manaka et~al.}(1997)\citenamefont{Manaka and Yamada}}]{Manaka1997}
\bibinfo{author}{\bibfnamefont{H.}~\bibnamefont{Manaka}} \bibnamefont{and}
  \bibinfo{author}{\bibfnamefont{I.}~\bibnamefont{Yamada}},
  \bibinfo{journal}{J. Phys. Soc. Jpn.} \textbf{\bibinfo{volume}{66}},
  \bibinfo{pages}{1908} (\bibinfo{year}{1997}).

\bibitem[{\citenamefont{Onodera et~al.}(1987)\citenamefont{Onodera, nakai,
  Kumitomi, Pringle, Smith, Nicklow, moon, Amita, Yamamoto, Kawano
  et~al.}}]{Onodera1987}
\bibinfo{author}{\bibfnamefont{A.}~\bibnamefont{Onodera}},
  \bibinfo{author}{\bibfnamefont{Y.}~\bibnamefont{Nakai}},
  \bibinfo{author}{\bibfnamefont{N.}~\bibnamefont{Kumitomi}},
  \bibinfo{author}{\bibfnamefont{O.~A.} \bibnamefont{Pringle}},
  \bibinfo{author}{\bibfnamefont{H.~G.} \bibnamefont{Smith}},
  \bibinfo{author}{\bibfnamefont{R.~M.} \bibnamefont{Nicklow}},
  \bibinfo{author}{\bibfnamefont{R.~M.} \bibnamefont{Moon}},
  \bibinfo{author}{\bibfnamefont{F.}~\bibnamefont{Amita}},
  \bibinfo{author}{\bibfnamefont{N.}~\bibnamefont{Yamamoto}},
  \bibinfo{author}{\bibfnamefont{S.}~\bibnamefont{Kawano}},
  \bibinfo{author}{\bibfnamefont{N.}~\bibnamefont{Achiwa}},
  \bibnamefont{and}
  \bibinfo{author}{\bibfnamefont{Y.}~\bibnamefont{Endoh}},
  \bibinfo{journal}{Jpn. J. Appl. Phys., Part 1}
  \textbf{\bibinfo{volume}{26}}, \bibinfo{pages}{152} (\bibinfo{year}{1987}).

\bibitem[{\citenamefont{Zaliznyak et~al.}(1998)\citenamefont{Zaliznyak, Dender,
  Broholm, and Reich}}]{Zaliznyak1998}
\bibinfo{author}{\bibfnamefont{I.~A.} \bibnamefont{Zaliznyak}},
  \bibinfo{author}{\bibfnamefont{D.~C.} \bibnamefont{Dender}},
  \bibinfo{author}{\bibfnamefont{C.}~\bibnamefont{Broholm}}, \bibnamefont{and}
  \bibinfo{author}{\bibfnamefont{D.~H.} \bibnamefont{Reich}},
  \bibinfo{journal}{Phys. Rev. B} \textbf{\bibinfo{volume}{57}},
  \bibinfo{pages}{5200} (\bibinfo{year}{1998}).

\bibitem[{\citenamefont{Zheludev et~al.}(2008)\citenamefont{Zheludev, Garlea,
  Regnault, Manaka, Tsvelik, and Chung}}]{Zheludev2008}
\bibinfo{author}{\bibfnamefont{A.}~\bibnamefont{Zheludev}},
  \bibinfo{author}{\bibfnamefont{V.~O.} \bibnamefont{Garlea}},
  \bibinfo{author}{\bibfnamefont{L.-P.} \bibnamefont{Regnault}},
  \bibinfo{author}{\bibfnamefont{H.}~\bibnamefont{Manaka}},
  \bibinfo{author}{\bibfnamefont{A.}~\bibnamefont{Tsvelik}}, \bibnamefont{and}
  \bibinfo{author}{\bibfnamefont{J.-H.} \bibnamefont{Chung}},
  \bibinfo{journal}{Phys. Rev. Lett.} \textbf{\bibinfo{volume}{100}},
  \bibinfo{pages}{157204} (\bibinfo{year}{2008}).

\bibitem[{\citenamefont{Xu et~al.}(2007)\citenamefont{Xu, Broholm, Soh, Aeppli,
  DiTusa, Chen, Kenzelmann, Frost, Ito, Oka, and Takagi}}]{Xu2007}
\bibinfo{author}{\bibfnamefont{G.}~\bibnamefont{Xu}},
  \bibinfo{author}{\bibfnamefont{C.}~\bibnamefont{Broholm}},
  \bibinfo{author}{\bibfnamefont{Y.-A.} \bibnamefont{Soh}},
  \bibinfo{author}{\bibfnamefont{G.}~\bibnamefont{Aeppli}},
  \bibinfo{author}{\bibfnamefont{J.~F.} \bibnamefont{DiTusa}},
  \bibinfo{author}{\bibfnamefont{Y.} \bibnamefont{Chen}},
  \bibinfo{author}{\bibfnamefont{M.} \bibnamefont{Kenzelmann}},
  \bibinfo{author}{\bibfnamefont{C.~D.} \bibnamefont{Frost}},
  \bibinfo{author}{\bibfnamefont{T.}~\bibnamefont{Ito}},
  \bibinfo{author}{\bibfnamefont{K.}~\bibnamefont{Oka}}, \bibnamefont{and}
  \bibinfo{author}{\bibfnamefont{H.}~\bibnamefont{Takagi}},
  \bibinfo{journal}{Science} \textbf{\bibinfo{volume}{317}},
  \bibinfo{pages}{1049} (\bibinfo{year}{2007}).

\end{thebibliography}

\end{document}